\documentstyle[12pt]{article}

\newcommand{\be}{\begin{equation}}
\newcommand{\ee}{\end{equation}}
\newcommand{\bean}{\begin{eqnarray*}}
\newcommand{\eean}{\end{eqnarray*}}
\newcommand{\bea}{\begin{eqnarray}}
\newcommand{\eea}{\end{eqnarray}}

\newcommand{\sq}{\sqrt{|g|}}

\author{M. G. Ivanov\thanks{e-mail: mgi@mi.ras.ru}\\
{\small\em Moscow Institute of Physics and Technology,}\\
{\small\em  141~700, Dolgoprudnyi, Moscow Region, Russia }}
\title{Black holes with complex multi-string configurations}
\date{November 5, 2001}

\begin{document}

\maketitle

\abstract{
  New exact solutions of Einstein equations which describe black hole with
 radial cosmic strings are constructed in the paper.
  The case of infinitely thin strings and the case of delocalized
 strings are considered.
  The case of delocalized strings allows generalization to 
 dimensions greater than 4.
 }

\section{Introduction}

  4D static charged black hole ${\bf M}_{\rm BH}$ has a form of
 warped product with warp factor $r^2$ of
 2D charged black hole ${\bf B}$ with metric
\be
  ds^2_{\bf B}=
   -\left(1-\frac{2M}{r}+\frac{Q^2}{r^2}\right)dt^2
   +\frac{dr^2}{1-\frac{2M}{r}+\frac{Q^2}{r^2}},
  ~~~-\infty<t<+\infty,~~0<r<+\infty,
\label{BH2D}
\ee
 and 2D unit sphere $S^2$ with metric
$$
  ds^2_{S^2}=\sin^2\theta~d\phi^2+d\theta^2,
  ~~~0\leq\theta\leq\pi,~~0\leq\phi<2\pi,
$$
 coordinate $\phi$ is periodic with period $2\pi$,
 i.e.
$$
  ds^2=ds^2_{\bf B}+r^2~ds^2_{S^2}.
$$

  Exact solution for a static black hole pierced by a single
 straight infinitely thin cosmic string ${\bf M}_{\rm cone}$
 \cite{AFV} has a similar form,
 but instead of sphere $S^2$ one has $S^2_{\rm cone}$ with metric
\be
  ds^2_{S^2_{\rm cone}}=\sin^2\theta~d\phi^2+d\theta^2,
  ~~~0\leq\theta\leq\pi,~~0\leq\phi<2\pi(1-\alpha),
\label{S2cone}
\ee
 coordinate $\phi$ is periodic with period $2\pi(1-\alpha)$
 (Riemannian geometry in space-time with conical defects
 was considered in \cite{FS}).
  String tension is proportional to angle deficit $\alpha$.
  Similarly, if one glues two equal spherical triangles
 along the coinciding edges to build manifold $S^2_{\rm triangle}$,
 then warped product with warp factor $r^2$ of
 ${\bf B}$ (\ref{BH2D}) and $S^2_{\rm triangle}$
 corresponds to configuration ${\bf M}_{\rm triangle}$
 of black hole with three radial cosmic strings \cite{DoCh:92}.
  Exact solution ${\bf M}_{\rm polyhedron}$,
 which describe black hole with $N_s$ radial strings
 directed along symmetry axes of a regular polyhedron \cite{FF},
 is similarly a warped product with warp factor $r^2$ of
 ${\bf B}$ (\ref{BH2D}) and $S^2_{\rm cone}[N_s]$.
  Manifold $S^2_{\rm cone}[N_s]$ can be constructed by gluing
 of $2(N_s-2)$ isometric spherical triangles.

  Metric of all the solutions above in the vicinity of string has 
 the form $ds^2=ds^2_{\bf B}+r^2~ds^2_{S^2_{\rm cone}}$,
 $0\leq\theta\leq\theta_0$.
  Locally the metric outside the strings is black hole metric.
  So, all solutions of this sort are build from standard blocks.

  The motivation of this work comes from the paper of
 Frolov and Fursaev \cite{FF}.
  They ask the question whether one  can
 obtain a closed 2D surface with conical
 singularities by gluing a certain set of
 different spherical triangles to find a
 general multi-string configuration attached
 to a black hole.
  We construct such solutions and their generalization in the limit
 of infinite number of infinitely light strings (continuous limit).

\section{Notation}
  $D$-dimensional space-time, i.e. (pseudo)Riemannian
 manifold ${\bf M}$ with metric 
 $ds^2=g_{MN}dX^MdX^N$, $M,N,\dots=0,\dots,D-1$
 is considered.

  For differential forms $A$ and $B$ with components
 $A_{M_1\dots M_q}$ and $B_{N_1\dots N_p}$
 the following tensor is introduced
$$
 (A,B)_{M_{k+1}\dots M_q~N_{k+1}\dots N_p}^{(k)}=\frac{1}{k!}~
         g^{M_1N_1}\dots g^{M_kN_k}
         A_{M_1\dots M_q}B_{N_1\dots N_p}.
$$
  Index $(k)$ indicates the number of indices to contract.
  If it can not lead to ambiguity, $(k)$ is skipped.

  For differential form of power $q$ it is
 convenient to introduce norm
 $\|A\|^2=(A,A)^{(q)}$,
 Hodge duality operation $*A=(\Omega,A)^{(q)}$,
 where $\Omega=\sq~d^DX$ is form of volume, and
 operation $\delta=*^{-1}d*$.
  Here and below $g=\det(g_{MN})$.

\section{Multi-string configuration}
  The action of the considered system is \cite{GSW,BN,VV}
\be
  S=\int_{\bf M}d^4X\sqrt{-g}
    \left(\frac{R}2-\frac12\|F\|^2\right)
   -\sum_s\mu_s\int_{{\bf V}_s} d^2\xi_s\sqrt{-\gamma_s},
\label{act}
\ee
 where index $s$ numerates strings, $\mu_s$ is string tension,
 $\xi^i_s$, $i=1,2$ are world-sheet coordinates,
 $(\gamma_s)_{ij}$ is metric induced at world-sheet ${\bf V}_s$.
  2-form $F=dA$ is electromagnetic field.

  Similar to solutions discussed in Introduction, let us consider
 space ${\bf M}_{\rm complex}$, which is warped product with warp
 factor $r^2$ of 2D charged black hole ${\bf B}$ (\ref{BH2D}) and 2D space
 $S^2_{\rm complex}$.
  Space $S^2_{\rm complex}$ is built by gluing of spherical
 triangles cut off a unit sphere.
  One can identify any two equal edges of triangles considered.
  Space ${\bf M}_{\rm complex}$ outside strings is locally
 isometric with black hole space ${\bf M}_{\rm BH}$.
  In the vicinities of strings, which correspond to vertices
 of triangles, ${\bf M}_{\rm complex}$ isometric with
 ${\bf M}_{\rm cone}$ for certain value of angle deficit $\alpha$.
  String tension $\mu_s$ and angle deficit $\alpha_s$ are related
 as $\mu_s=2\pi\alpha_s$.
  Manifolds ${\bf M}_{\rm cone}$ and ${\bf M}_{\rm BH}$ are
 solutions of the model (\ref{act}).
  Model (\ref{act}) is local, so ${\bf M}_{\rm complex}$ is
 also a solution of the same model.

  Solution ${\bf M}_{\rm complex}$ may bear no global symmetry, but
 in the vicinity of every string it is invariant under O(2) rotations
 about the string.

  One can describe the process of building $S^2_{\rm complex}$
 in terms of gluing of conical defects.
  In this case there are some non-trivial consistency conditions
 written in terms of angle deficits.
  In this paper we are not looking for such conditions.
  We consider the process of smooth gluing of spherical triangles
 cut off a unit sphere.
  In this case the only consistency conditions is equal lengths
 of edges we glue.
  The procedure described below is just a formalization of the
 simple idea above.

  One can use the following algorithm of building
 closed space $S^2_{\rm complex}$.
  Let us introduce 
 closed 2D simplicial complex $c^2$ ($\partial c^2=0$),
 such that
 every point of the complex belongs to finite number of simplexes,
 for non-equal 2D simplexes
 $\sigma^2_1,\sigma^2_2,\sigma^2_3\in c^2$,
 $\sigma^2_1\cap\sigma^2_2\cap\sigma^2_3$ is point or empty set,
 and length function $L$, which maps all 1D simplexes of $c^2$
 to segment $(0,\pi/2]$.
  If $\partial\sigma^2=\sigma^1_1+\sigma^1_2+\sigma^1_3$,
 where $\sigma^2\in c^2$ is 2D simplex, then 
 $L(-\sigma^1_1)=L(\sigma^1_1)<L(\sigma^1_2)+L(\sigma^1_3)$.
  It is possible to continuously map every 2D simplex $\sigma^2\in c^2$
 with boundary $\sigma^1_1+\sigma^1_2+\sigma^1_3$ to spherical 
 triangle of unit sphere with edges
 $L(\sigma^1_1)$, $L(\sigma^1_2)$, $L(\sigma^1_3)$.
  To make the algorithm unambiguous one have to choose spherical
 triangles, which area does not exceed $\pi/2$.

  One can consider similar gluing procedure with spherical
 triangles cut off sphere of any radius $R\in[1,\infty]$
 ($R=\infty$ corresponds to planar triangles).
  For any vertex $v$ of the simplicial complex above one can
 calculate angle deficit
 $\alpha(v,R)=\frac1{2\pi}(2\pi-\phi_{v1}(R)-\dots-\phi_{vk_v}(R))$,
 where $\phi_{v1}(R),\dots,\phi_{vk_v}(R)$ are adjacent to $v$ angles of
 spherical triangles adjacent to $v$.
 $\frac{d}{dR}\phi_{vn}(R)>0$, so $\frac{d}{dR}\alpha(v,R)<0$.
  So, for finite simplicial complex,
 if $\min\alpha(v,\infty)>0$ ({\em finite convex complex}),
 then there exists radius $R_0\in[1,\infty)$,
 such that for any $v$ and any $R\in(R_0,\infty)$
 angle deficit is positive $\alpha(v,R)>0$.
  Obviously $\alpha(v,R)=\tilde\alpha(v,1)$, where
 $\tilde\alpha(v,1)$ is angle deficit calculated with length function
 $\tilde L=\frac1RL$ instead of $L$.
  It allows as to rescale any finite convex complex to guarantee
 positive $\alpha(v,1)$ and positive string tensions.
  A simple example of finite convex complex is any convex polyhedron
 (to convert it into simplicial complex one have to add some extra
  edges to split faces into triangles).
 
\section{Continuous limit}
  Let us consider the following action
\be
  S=\int_{\bf M}d^4X\sqrt{-g}
    \left(\frac{R}2-\frac12\|F\|^2-\|J\|\right),
\label{act4DJ}
\ee
 where 2-form $F=dA$ is electromagnetic field,
 and 2-form $J=d\varphi^1\wedge d\varphi^2$ is string (membrane) field
 (see \cite{hos1}--\cite{MGI3}).
  Norm $\|J\|$ represents density of string matter.
  By integration of $J$ over 2D surface ${\bf U}$ one can find the
 total tension $\mu_{\bf U}$ of strings (string flux),
 which intersect the surface
\be
  \mu_{\bf U}=\int_{\bf U}J.
\label{UJ}
\ee

  One can consider action (\ref{act4DJ}) as continuous limit of
 thin string action (\ref{act}) discussed in previous section.
  Vice versa thin string action (\ref{act}) corresponds to
 singular limit of action (\ref{act4DJ}).
  So, the formulae of this section under appropriate interpretation
 are applicable to thin strings case too.

  By variation of fields $g_{MN}$, $A_M$ and $\varphi^\alpha$
 ($\alpha=1,2$) one can find equations of motion
\bea
  R^M_N-\frac12\delta^M_NR=
   (F,F)^M_N-\frac12\|F\|^2\delta^M_N
  +\frac1{\|J\|}~(J,J)^M_N-\|J\|\delta^M_N,
  \label{eomM}
  \\\nonumber
  \delta F=0,~~~\delta\frac{J}{\|J\|}=0.
\eea
  Let space-time ${\bf M}$ is warped product with warp factor $r^2$
 of 2D black hole ${\bf B}$ (\ref{BH2D}) and an arbitrary
 2D Riemannian manifold
 ${\bf S}$ with metric $ds^2_{\bf S}=e^{2f(x,y)}(dx^2+dy^2)$.

  The following fields are solution of equations of motion (\ref{eomM})
\bea
  ds^2=-\left(1-\frac{2M}{r}+\frac{Q^2}{r^2}\right)dt^2
       +\frac{dr^2}{1-\frac{2M}{r}+\frac{Q^2}{r^2}}
       +r^2e^{2f(x,y)}~(dx^2+dy^2),
  \label{cont}
  \\\nonumber
  F=\sqrt2~\frac{Q}{r^2}~dt\wedge dr,~~~J=({\cal K}-1)~\Omega_{\bf S},
\eea
 where $\Omega_{\bf S}=e^{2f(x,y)}~dx\wedge dy$ is 2-form of volume 
 at ${\bf S}$, and
\be
  {\cal K}=-e^{-2f(x,y)}\triangle f(x,y)
\ee
 is Riemannian curvature of ${\bf S}$
 ($\triangle=\partial_x^2+\partial_y^2$).
  In definition of norm $\|J\|$ one have to choose branch of square root,
 which corresponds to $\|J\|=\frac{{\cal K}-1}{r^2}$, so
 the case ${\cal K}<1$ corresponds to negative string tension.

  For compact ${\bf S}$ by integrating of $J$ defined by (\ref{cont})
 the total string tension is (\ref{UJ})
\be
  \mu_{\bf S}=2\pi\chi({\bf S})-A({\bf S}),
\ee
 where $\chi({\bf S})$ is Euler characteristic of ${\bf S}$,
 and $A({\bf S})>0$ is area of ${\bf S}$.
  If one require string tension to be positive, then
 $\mu_{\bf S}>0$, and $\chi({\bf S})>0$ is necessary
 condition on topology of ${\bf S}$.

\section{Multidimensional continuous case}
  In multidimensional case straightforward generalization of 
 multi-string solutions is not possible, but one can generalize
 spherically symmetric version of continuous solution (\ref{cont}).
  Let us consider the action
\be
  S=\int_{\bf M}d^DX\sqrt{-g}
    \left(\frac{R}2-\frac12\|F\|^2-\|J\|\right),
\ee
 where 2-form $F=dA$ is electromagnetic field,
 and $(D-2)$-form $J=d\varphi^1\wedge\dots\wedge d\varphi^{D-2}$
 is string field.
  By variation of fields $g_{MN}$, $A_M$ and $\varphi^\alpha$
 ($\alpha=1,\dots,D-2$) one can find equations of motion, which
 have the form identical to (\ref{eomM}).
  
  The following fields are solution of equations of motion (\ref{eomM})
\bea
  ds^2=-\left(1-\frac{2(M+\mu r)}{r^{D-3}}+\frac{Q^2}{r^{2(D-3)}}\right)dt^2
       +\frac{dr^2}{1-\frac{2(M+\mu r)}{r^{D-3}}+\frac{Q^2}{r^{2(D-3)}}}
       +r^2 g^{\bf S}_{ij}dx^idx^j
  \label{contD}
  \\\nonumber
  F=\sqrt{(D-3)(D-2)}~\frac{Q}{r^{D-2}}~dt\wedge dr,
  ~~~J=(D-2)~\mu~\Omega_{\bf S},
\eea
 where $x^i$ ($i,j=1\dots D-2$) are coordinates at
 $(D-2)$-dimensional Riemannian manifold ${\bf S}$
 with metric $ds^2_{\bf S}=g^{\bf S}_{ij}dx^idx^j$,
 which satisfies the condition
 (Einstein equations at ${\bf S}$)
\be
  R^{\bf S}_{ij}=(D-3)~g^{\bf S}_{ij}
\label{RgS}
\ee
 (e.g. unit sphere $S^{D-2}$).
 $\Omega_{\bf S}$ is volume form at ${\bf S}$.
  In definition of norm $\|J\|$ one have to choose branch of square root,
 which corresponds to $\|J\|=(D-2)~\frac{\mu}{r^{D-2}}$.

\section{Uniqueness of 4D case}
  The effective mass of solution (\ref{contD})
 at large $r$ is linear in $r$.
  One can expect this feature of solution a priori, because string
 energy is proportional to string length and radial strings
 pierce every surface $r=const>r_0$.
  Newtonian gravitational potential is $\frac{M+\mu r}{r^{D-3}}$, so only
 in 4D case one can express radial uniform string distribution 
 in terms of deformation of surface ${\bf S}$
 (this surface up to scale factor is isometric to horizon).

  In 4D case surface ${\bf S}$ is 2D.
  2D Riemannian tensor has only one independent component $R_{1212}$,
 it forces manifold ${\bf S}$ in (\ref{contD}) to be a unit sphere (locally),
 but in solution (\ref{cont}) one is able to choose any manifold to be
 the horizon.
  So, using solutions (\ref{cont}), (\ref{contD}) it is possible
 to choose any horizon geometry in 2D case.
  If $D>4$ horizon geometry is restricted by equation (\ref{RgS}).
  In 4D case 2D horizon geometry defines the density of string matter
 in the bulk.
  If $D>4$ string matter density is controlled by a single parameter $\mu$.

\section{Discussion}
  The purpose of this work was to find exact static solutions which
 describe black hole with radial strings.
  In 4D case the solutions were built using thin strings or 
 string field (\ref{cont}).
  4D case appears to be very special
 (see discussion in the previous section).

  In multidimensional cases ($D>4$) solutions (\ref{contD})
 were found only in terms of string field.
  String matter density $\|J\|=(D-2)~\frac{\mu}{r^{D-2}}$ is
 constant along $r=const$ surfaces, so solutions of
 form (\ref{contD}) ($D>4$) do not describe thin strings.

  It is an interesting problem to find a static black hole solution
 with radial thin strings in multidimentional case ($D>4$) or
 to prove absence of solutions of this sort. 

  Another problem is to study static black p-brane solutions
 with radial membranes of different dimensions.

  The author is grateful to I.V. Volovich, M.O. Katanaev,
 V.P. Frolov and D.V. Fursaev.
  The work was partially supported by grant RFFI 99-01-00866.

\end{document}